\documentstyle[12pt]{article}

\input{epsf}

\begin{document}
\rightline{RU-96-93} 
\newcommand{\pho}{\tilde{\gamma}}
\newcommand{\gl}{\tilde{g}}
\newcommand{\sneu}{\tilde{\nu}}
\newcommand{\sq}{\tilde{q}}
\newcommand{\se}{\tilde{e}}
\newcommand{\eff}{\epsilon_{6/4}}
\newcommand{\ch}{\chi^{\pm}}
\newcommand{\neut}{\chi^{0}}
\newcommand{\gsi}{\,\raisebox{-0.13cm}{$\stackrel{\textstyle>}
{\textstyle\sim}$}\,}
\newcommand{\lsi}{\,\raisebox{-0.13cm}{$\stackrel{\textstyle<}
{\textstyle\sim}$}\,}
\baselineskip=18pt \vskip 0.7in 
\begin{center} {\bf \large
Phenomenology of ``inos'' in the Light Gaugino Scenario, \\
and Possible Evidence for a $\sim 53$ GeV Chargino} \\ 
\vspace*{0.9in} 
{\large Glennys R. Farrar}\footnote{Invited talk at ICHEP96, Warsaw,
July, 1996.  Based on contributed paper pa11-048 and RU-96-71
(hep-ph/9608387).  Research supported in part by NSF-PHY-94-23002} \\ 
\vspace{.1in} 
{\it Department of Physics and Astronomy \\ Rutgers
University, Piscataway, NJ 08855, USA}\\ 
\end{center} 
\vspace*{0.2in}
\vskip 0.3in 
{\bf Abstract:} The tree-level-massless gaugino scenario predicts that the lighter
chargino mass is less than $m_W$ and that gluino and lightest
neutralino masses are $ \lsi 1$ GeV.  In this case the dominant decay
mode of charginos and non-LSP neutralinos is generically to three
jets.  The excess of "$4j$" events with total invariant mass
$\sim 105$ GeV observed in LEP running at 130-136 GeV is noted to be
consistent with pair production of $\sim 53$ GeV charginos.  
Data at 161 and 172 GeV from Fall, 1996, cannot conclusively test this
hypothesis (because cuts to eliminate $W^+W^-$ background reduce the
efficiency significantly) but is suggestive that the signal persists.

\thispagestyle{empty} 
\newpage 
\addtocounter{page}{-1}
\newpage

Some supersymmetry (SUSY) breaking scenarios produce negligible
tree-level gaugino masses and scalar trilinear couplings
($M_1=M_2=M_3=A=0$).  This has several attractive theoretical
consequences such as the absence of the ``SUSY CP problem" and
avoidance of certain cosmological problems\cite{f99101}.  Although
massless at tree level, gauginos get calculable masses through
radiative corrections from electroweak (gaugino/higgsino-Higgs/gauge
boson) and top-stop loops. Evaluating these within the constrained
parameter space leads to a gluino mass range $m_{\tilde{g}}\sim
\frac{1}{10} - \frac{1}{2}$ GeV and photino mass range
$m_{\tilde{\gamma}} \sim \frac{1}{10} - 1 \frac{1}{2}$
GeV\cite{f99101}.  The chargino and other neutralino masses are, at
tree level, functions only of $\mu$ and $tan \beta$.  In particular, 
\begin{equation}
2 M_{\ch}^2 = \mu^2 + 2 m_W^2 \pm \sqrt{\mu^4 + 4 m_W^4 cos^2
2 \beta + 4 m_W^2 \mu^2}, 
\label{mch}
\end{equation}
so one chargino is lighter, and the other heavier, than $m_W$.  The
photino is an attractive dark matter candidate, with a correct abundance
for parameters in the predicted ranges\cite{f:100}. 

Due to the non-negligible mass of the photino compared to the
lightest gluino-containing hadron, prompt photinos\cite{f:24} are not
a useful signature for the light gluinos and the energy they
carry\cite{f:51}.  Gluino masses less than about $ \frac{1}{2}$ GeV are
largely unconstrained\cite{f:95}.  [The recent claims\cite{murfod}
that LEP $Z^0 \rightarrow 4j$ data can be used to exclude light
gluinos are premature.  The statistical power of the data is indeed
sufficient, however the relevant angular distributions are sensitive
to jet definition and higher order effects such as 5-jet production, so
systematic uncertainties dominate and no conclusions can be drawn
until such effects can be controlled\footnote{C.f., J. W. Gary, CTEQ
Workshop on QCD, FNAL, Nov. 1996.}]  The lifetime of the gluon-gluino
bound state ($R^0$) is predicted to be $10^{-5}-10^{-10}$
sec\cite{f99101}.  Proposals for direct searches for hadrons containing
gluinos, via their decays in $K^0$ beams and otherwise, are given in
Ref. \cite{f:104,nussinov}.  

For the purposes of detecting squarks and charginos, the crucial
phenomenological difference arising when the gluino is light rather
than heavy as is usually assumed, is that the gluino is long-lived.
Therefore it makes a jet rather than missing energy\cite{f:51}.
Squarks decay to gluino and quark, thus generating two jets with
negligible missing energy.  QCD background makes it impossible, with
present jet resolution, to search at hadron colliders in the dijet
channel for masses lower than about 200 GeV; a search for equal-mass
dijet pairs as suggested in \cite{f:105} has not yet been completed.
At present, the best squark mass limits come from the hadronic width
of the $Z^0$ and are only $\sim 50-60$ GeV\cite{sqlim,f:105}.

If the gluino is light, the lighter chargino and the three
heavier neutralinos will usually decay to $q \bar{q'} \gl$, via a
virtual or real squark\footnote{For $M_{\sq} \sim m_W$, decay via
virtual squark is a factor $\frac{2 \cdot 8 \alpha_3}{(2 \cdot 3 + 3)
\alpha_2} \sim 5$ larger than via a virtual $W$, summing over the two
generations of light quarks and three generations of leptons, and a
factor $8\frac{\alpha_3}{\alpha_2} \sim 25$ larger than via virtual
slepton or sneutrino.  When $tan \beta \approx 1$, $\neut_3 \approx
\tilde{h}_U + \tilde{h}_D$ and has very little $\tilde{z}$ component;
its dominant decay mode may be $\neut_3 \rightarrow \gamma \pho$,
through a stop-top loop.}, unless all flavors of squarks are much more
massive than sneutrinos, sleptons and $W$'s.  The heavier chargino
generally has the kinematically allowed two-body decay $\ch_2
\rightarrow W^{\pm} + \pho$.    

Although $\ch_1 \rightarrow q \bar{q'} \gl$ gives a 3-parton final
state, it may be reconstructed as $2j$, particularly when one jet is
much softer than the others.  Then, the hadrons of the soft jet have
low invariant mass with respect to hadrons of the other jets and the
jet finding algorithm can interpret the multijet system as $2j$. 
The likelihood of this increases when $M_{\ch} - M_{\sq} << M_{\ch}$,
as is illustrated by the cascade chain $\ch \rightarrow \sq 
+ q'~, \sq \rightarrow \gl q$, for which $E(q')\approx
M_{\ch}-M_{\sq}$ and $E(\gl) = E(q) = M_{\sq}/2$, in the $\ch$ rest
frame.  Below we will use $P_{6},~ P_{5},~P_4$ to denote the
probabilities that $\ch \chi^{\mp}$ decay produces events which are
designated $6j,~5j,~4j$ in the experimental analysis. Since these
depend strongly on $M_{\ch}-M_{\sq}$, a limit on, say, $P_6/P_4$
implies a limit on the mass splitting between chargino and intermediate
squark.   

In the LEP 130 and 136 GeV run, ALEPH observed 16 $4j$ events when 8.6
were expected \cite{aleph:4j}.  Their jet-reconstruction algorithm
explicitly merged $5j$ to $4j$ (i.e., $P_5 \equiv 0$), and ignored the
small number of clear $6j$ events\footnote{M. Schmitt, private
communication.}.  Nine events were observed in the total-dijet-mass
range 102-108 GeV when 0.8 event was expected.   Most of the events in
the peak region are not characteristic of the SM expectation in
their kinematic distribution, and they have a dijet charge-difference
$\Delta Q$ larger than expected in the SM and well-described by a
parton-level $\Delta Q = 2$ as predicted for $\ch \chi^{\mp}$ events.
A statistical fluctuation or experimental reconstruction artifact
would not readily account for such deviations from SM event
characteristics.    

The other three LEP experiments, using the ALEPH analysis procedure,
found 6 events in the 102-108 GeV bin, when 2.6 were expected.
Including events from 19 pb$^{-1}$ of data at $E_{cm} = 161$ and 171
GeV, ALEPH finds a total of 18 events in the peak region when 3.1 are
expected\footnote{F. Ragusa, LEPC Nov. 19, 1996}.  The other
experiments have reported neither a significant signal nor upper
limit, so far\footnote{C.f., joint LEP seminar, CERN, Oct. 8, 1996.}.   

It should be emphasized that there is a substantial uncertainty in the
expected rate of $\ge 4$ jet events due to the renormalization scale
sensitivity of the tree-level cross sections which have been used.
For instance, the observed rate of $Z^0 \rightarrow 4j~ (5j)$ is a
factor 3 (5) higher than predicted taking the scale $\mu =
m_Z$\cite{opal:5j}.  Thus the important feature of the LEP $4j$
anomaly is the {\it peaking} in total invariant mass and the anomalous
properties of the correct fraction of peak region events. 

In order to quote a cross section corresponding to the observed rate
of anomalous events it is necessary to make an assumption as to the
source of the signal.  Taking it to be due to pair production of equal
mass particles which decay to two jets\footnote{E.g., $h~A$, although
that is unlikely to be the origin of the excess events, since the predicted
cross section is 0.49 pb for $M(h) = M(A) = 53$ GeV and there is no
observed excess of $b \bar{b}$'s in the events.} implies a cross
section at 130-136 GeV of $3.1~\pm~1.7$ pb using ALEPH alone, or $1.2 ~\pm
~0.4$ pb averaging all LEP experiments\cite{mattig}.  Averaging the
161 GeV data available at Warsaw, Mattig quotes a 95\% cl upper limit
of 0.85 pb assuming the same efficiency as at 130-136 GeV.  Extrapolating
1.2 pb from 130-136 GeV with $1/s$ gives 0.8 pb.

Pair production of $\sim 53$ GeV charginos could give rise to ``$4j$''
events with total dijet mass of 105 GeV at the observed rate and with
the observed $\Delta Q$ values.  Let us denote by $\epsilon_{6/4}$ the
ratio of probabilities that a $\ch \chi^{\mp} \rightarrow 6j$ or a
$h A \rightarrow 4j$ event is accepted by the ALEPH cuts.  Like the
$P_i$'s defined above, $\epsilon_{6/4}$ depends on $M_{\sq}$ as well
as $E_{cm}$.  A careful Monte Carlo calculation is needed to determine 
these quantities.  The two (hard) gluino jets contain
$R^0$'s\footnote{Refs. \cite{nussinov,deR_P} estimate their $x_F$.}
which may decay before reaching the calorimeter (typically, to $\pi^+
\pi^- \pho$\cite{f:104}).  Due to experimental imprecision in energy
measurements, the ALEPH $4j$ analysis rescales the momenta and
energies of jets to obtain an improvement in precision by enforcing
overall energy momentum conservation.  The directions of the jets are
assumed to be accurately measured, and the jets taken to be massless.
Energy-momentum conservation provides 4 equations, so up to 4 jets can
be independently rescaled.  If an event actually contains 6 jets,
possibly with some energy and directional imprecision due to $R^0$
decays, the rescaling procedure distorts the dijet invariant masses.
Detailed Monte Carlo study is needed to estimate the $P_i$'s, $\eff$
and predict the dijet mass-sum and mass-difference distributions and
other observables. 

The chargino production cross section depends on $\mu,~tan \beta$ and
$M_{\sneu_e}$.  $M_{\ch_1}= 53$ GeV requires a relation between
$\mu$ and $tan \beta$, eq. (\ref{mch}).  Imposing $M_{\neut_2} \ge 38$
GeV (see below), causes $[\mu,tan\beta]$ to range between $[45,1.6]$
and $[70,1] $\footnote{We take $tan \beta \ge 1$ without loss of
generality because in the absence of tree-level gaugino masses and
scalar trilinear couplings, the chargino and neutralino spectrum is
unchanged by $tan \beta \rightarrow 1/(tan \beta)$; only the roles of
the higgsinos, $h_U$ and $h_D$, are interchanged in the eigenstates.}.
In this constrained parameter space the chargino contains comparable
higgsino and wino components.   The main uncertainty in the cross
section comes from its sensitivity to the electron sneutrino mass,
$M_{\sneu_e}$. For $M_{\sneu_e} = 60$ GeV, $tan \beta = 1.4$ and $\mu
= 56$ GeV, one finds\cite{mlm:xsecn} $\sigma(\chi^+ \chi^-) = 3.6$ pb
at $E = 130-136$ GeV, and 2.1 (1.9) pb at $E=161~(190)$ GeV.  For
$M_{\sneu_e} = 50$ GeV the lowest $\sigma_{\chi^+_1 \chi^-_1}(130-136 {\rm
GeV}) = 2.4$ pb, while for $M_{\sneu_e}>> M_{\ch_1}$ the 130-136 GeV cross
section could be as large as 14 pb, for $tan \beta = 1$ and $\mu = 68.4$ GeV.   

Comparing to the LEP average ($1.2 \pm 0.4$ pb, assuming $hA$ for the
efficiency) these predicted cross sections are compatible with
$br(\chi^{\pm} \rightarrow 3j) = 1$ if $\eff \lsi \frac{1}{2}$.  If
$\eff > \frac{1}{2}$, a competing decay mode to reduce
$br(\chi^{\pm} \rightarrow 3j)$ would be indicated.  This could be
$\ch_1 \rightarrow l \sneu$, with the $\sneu$ decaying to lsp and
neutrino. The branching fraction for $\ch_1 \rightarrow l \sneu$ is a
very sensitive function of the sneutrino mass and also depends on the
squark mass and the number of light sneutrinos.  The mass splitting
$M(\ch) - M(\sneu)$ must be $\lsi 1$ GeV in order not to excessively
reduce the branching fraction to the hadronic channel.  With such a
small splitting, the lepton energy is too low for lepton
identification so the events with one or both charginos decaying
leptonically would not be noticed.  However $\ch \rightarrow \tilde{l} 
\nu$ would give rise to a hard lepton and thus would show up in
conventional SUSY searches, so we can infer $M(\tilde{l}) \gsi M(\ch_1)$.

A stop lighter than the chargino, decaying through FCNC mixing to a
gluino and charm quark, may be excludable by ALEPH.  It would lead to
two hard $c$ jets in each event (plus two soft $b$ jets if $\ch_1
\rightarrow \tilde{t}_1 + b$ were allowed).  ALEPH has searched for
hard $b$ jets in their $4j$ sample.  They found only a single event
consistent with a displaced vertex, or excess lepton activity from the
$e$ or $\mu$, which would be expected in 20\% of hard $b$
decays\cite{aleph:4j}.  Although the detection efficiency of hard 
$c$'s and soft $b$'s is lower than for hard $b$'s, it may still be
large enough to exclude the dominant chargino decay being $\ch_1
\rightarrow \tilde{g} + c + b$.  This would either mean that
$M_{\tilde{t}_1} \gsi M_{\ch_1}$ or that the dominant decay of the 
stop is $\tilde{t}_1 \rightarrow \tilde{g} + u$.  Since FCNC mixing 
involving the third generation is poorly constrained, the latter
possibility should not be dismissed\footnote{If $M_{\tilde{t}_1} +
m_{\tilde{g}} < m_t $, $br(t \rightarrow \tilde{t}_1 + \tilde{g})
\approx \frac{1}{2}$ and $\tilde{t}_1 \rightarrow \chi^+_1 b$, $c
\tilde{g}$, or $u \gl$.  The rate of purely SM decays in $t \bar{t}$
events would be reduced, but present theoretical and experimental  
uncertainties in $\sigma(p \bar{p} \rightarrow t \bar{t})$ are too
large to exclude this, especially when the stop decays to $b$
quark plus multijets.  More promising is to compare the ratio of $t
\bar{t}$ events obtained with a single lepton versus dilepton tag.  I
am indebted to S. Lammel for a discussion of this point.}.   

We can hope that higher integrated luminosity will allow the 53 GeV
chargino hypothesis to be confirmed or excluded, however this will not
be easy.  In future runs at higher CM energy the jet systems from each
chargino decay will be better collimated and more readily separated
from one another, so the angular distribution of the jet systems can
be more cleanly determined than at lower energy.  It should $\sim 1 + cos^2
\theta$ because charginos are spin-1/2.  On the other hand, Monte
Carlo simulations\footnote{ (M. Schmitt, FNAL seminar, Nov. 22, 
1996.} show
that the resolution in dijet mass difference does not improve
significantly with energy even for genuine two-body decays of
pair-produced particles, due to reduced effectiveness of the
energy-momentum constraint at higher energy.  This will be an even
more severe problem for the $6j$ case at hand.  It is ironic that   
$W^+ W^-$ production is a non-trivial background at 161 GeV and
above, since $W$'s presumably decay with rather high probability to $>
2j$ (40 \% of $Z^0$ decays contain $\ge 3j$, with $y_{cut} = 0.01$)
and their cross section is much larger than that of the $4j$ signal.

\begin{figure}
\epsfxsize=\hsize
\epsffile{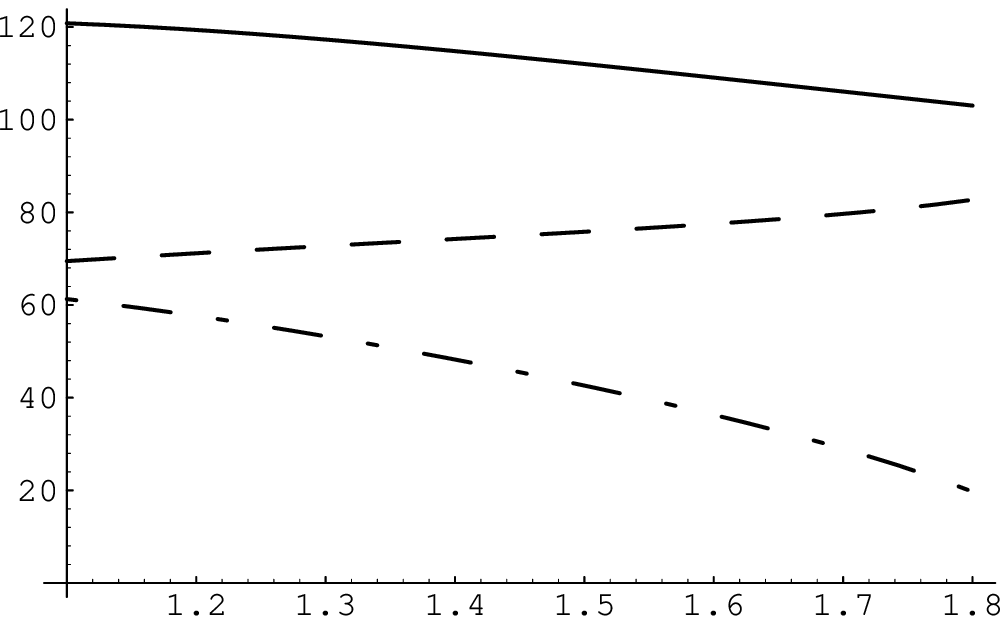}
\caption{Masses in GeV of the heavier chargino and second and third
neutralinos as a function of $tan \beta$, fixing the lighter chargino
mass to 53 GeV (solid, dash-dot, and dash curves, respectively).}  
\label{masses}
\end{figure}

Now let us turn to the neutralinos and heavier chargino whose masses
are shown in Fig. \ref{masses} for $M(\ch_1) = 53$ GeV.  $\neut_1
\neut_1$ and $\neut_1 \neut_2$ production is suppressed because
$\neut_1$ is practically pure photino.  Since the final state of
$\neut_2 \neut_2$ is $6j$, existing neutralino searches would not have
been sensitive to it, so the best limit comes from the hadronic width
of the $Z^0$, whose PDG value is $\Gamma_{had}(Z^0) = 1741 \pm 6$ MeV.
Requiring the $\neut_2 \neut_2$ contribution to be $< 10$ MeV implies
$M(\neut_2) \ge 38$ GeV, which in turn limits the $[\mu, tan \beta]$
range to $[45,1.6] - [70,1]$. For $M_{\neut_2} = 38$ GeV, $\neut_2$
has a large $\tilde{h}_D$ component ($\ch_2 = -0.86 \tilde{h}_D -0.39
\tilde{h}_U + 0.31 \tilde{z}^0$).  This leads to an excess in the $b
\bar{b} \gl$ final state in $\neut_2$ decay, but not enough to affect
$R_b$ significantly.  At 130-136 GeV, $\sigma(e^+ e^- \rightarrow \neut_2
\neut_2) < 0.5$ pb for typical parameter choices, although for
$M(\neut_2) \lsi 45$ GeV the cross section (including enhancement from
initial state radiation) is $\gsi 1$ pb.  Thus a small number of
events might be present in the LEP multi-jet samples.  Such events should
exhibit $\Delta Q$ consistent with zero.  Production of $\neut_2
\neut_3$ is at least an order of magnitude lower. 

Since $\ch_2 \rightarrow W \pho$, Drell-Yan production of $\ch_2
\ch_2$ and $\ch_2 \neut$ might be detectable in $p \bar{p}$
collisions via events with $3j + W + E_{miss}$, where the $3j$ system
has an invariant mass of $\ch_2$ or $\neut$ (see Fig. 1).  The
$6j$ signal would probably be difficult to discriminate from QCD
background.  For some range of parameters$^b$ $\neut_3 \rightarrow
\gamma \pho $ might lead to $\gamma E_{miss}$ at an interesting rate. 

To summarize:\\
$\bullet$ The tree-level-massless gaugino scenario predicts
$m(\chi^0_1) \lsi 1 $ GeV and $m(\chi^{\pm}_1) \lsi m_W$; the measured
$Z^0$ hadronic width requires $m(\chi^0_2) \gsi 38 $ GeV.  \\ 
$\bullet$ The generic final state for neutralinos and charginos in the
light gaugino scenario is three jets, unless $m(\sneu)$ or
$m(\tilde{l})$ is small enough that two-body leptonic decays are
allowed.  The heavier chargino has the characteristic 2-body decay
$\ch_2 \rightarrow W^{\pm} \pho$.\\ 
$\bullet$ Chargino production and decay may account for the rate and
characteristics of ``$4j$'' events seen at LEP, for chargino mass of
$\approx 53$ GeV.  No R-parity violation is needed. \\  
$\bullet$ If the present hint of charginos at 53 GeV disappears,
experiments should still search for charginos via a $6j$ signature
until light gluinos are ruled out.  At $E_{cm}=190$ GeV the cross
section for chargino pair production is greater than 1.4 pb for
$M_{\sneu_e} = 100$ GeV, even for the most pessimistic case of
degenerate charginos with mass $m_W$.  Above $WW$ threshold the $6j$
signal is difficult to discriminate from the background, however.  

{\bf Acknowledgements:}  I am indebted to many people for information
and helpful discussions, including M. Albrow, H. Baer, J. Berryhill,
J. Carr, J. Conway, J. Feng, H. Frisch, E. Gross, G. Kane, P. Janot,
M. Mangano, F. Richard, M. Schmitt, A. Tilquin, and S. L. Wu.   



\end{document}